\documentclass[aip,jap,preprint]{revtex4-2}

\usepackage{amsmath,amssymb,amsfonts,bm,mathtools}
\usepackage{graphicx}
\usepackage{xcolor}
\usepackage{hyperref}
\usepackage{physics}
\usepackage{booktabs}
\usepackage{array}
\usepackage{enumitem}
\usepackage{CJK}
\usepackage{chemfig,chemformula}

\newcommand{\etal}{\emph{et al.}}

\newcommand{\un}[1]{\ensuremath{\unskip\,\mathrm{#1}}}

\begin{document}
\begin{CJK}{UTF8}{gbsn}	
	\title{Impedance Matching and Absorption Enhancement in Helical Carbon Coil Microwave Absorbers via Tunable Anchoring Layer Thickness}
	
	\author{Weihua Mu (牟维华)}
	\affiliation{Wenzhou Key Laboratory of Biomaterials and Engineering, Wenzhou Institute, University of Chinese Academy of Sciences, Wenzhou, Zhejiang 325000, China}
	\affiliation{Zhejiang Key Laboratory of Soft Matter Biomedical Materials, Wenzhou Institute, University of Chinese Academy of Sciences, Wenzhou, Zhejiang 325000, China}
	\email{muwh@wiucas.ac.cn}
	
	\date{\today}
	
	\begin{abstract}
	Helical carbon coils exhibit unique three dimensional chiral architectures that can effectively interact with microwaves in the 2 to 18 GHz range. Inspired by recent experimental studies, we develop a coarse grained electrodynamic model for helical carbon coil arrays supported on quartz substrates, and examine their potential as microwave absorbers. Guided by the heuristic relation $f_{\mathrm{res}} = c / (n_{\mathrm{eff}} p)$, we compute absorption and reflection fractions for both bare helical carbon coil on substrate and helical carbon coil with anchoring layer on substrate configurations. The anchoring layer thickness is treated as a tunable parameter to improve absorption at selected frequencies. We find that the introduction of a carbon based anchoring layer of optimized thickness enhances impedance matching absorption. The parameters are chosen as representative values for demonstration purposes, and may vary within a certain range depending on preparation conditions. Our results serve to illustrate the potential of helical carbon coils as microwave absorbing devices, and to identify possible active tuning strategies. In the discussion section, we consider extensions of the present model to account for circular dichroism absorption, and examine the influence of actual interfacial reflection on microwave absorption and reflection. 
    \end{abstract}
	
	\maketitle
	
	\section{Introduction}
	
	Structured carbon materials have attracted great interest for microwave absorption applications, which is driven by the need for lightweight, broadband, and efficient absorbers~\cite{Wang2021JCIS,Hou2024Carbon}. Among these materials, helical carbon coils (HCCs) stand out as a promising chiral carbon material. They are amenable to large cale synthesis by chemical vapor deposition (CVD)~\cite{Motojima2004}. The HCCs offer high electrical conductivity, thermal and chemical stability, and unique mechanical resilience.
	These properties enable efficient attenuation of incident electromagnetic energy.
	
	Recent studies have expanded the microwave absorption capabilities of HCCs through structural and chemical modifications.
	Zuo et al. fabricated micro-mesopores on spiral carbon nanocoils and simultaneously doped them with oxygen to broaden the absorption bandwidth~\cite{Zuo2024AdvFuncMater}. Sun et al. demonstrated low cost broadband absorption by integrating helical carbon microcoils on toilet paper substrates~\cite{Sun2025CarbonHelix}. Particularly, Qi et al. investigated the circular dichroism of chiral helical carbon coils in the microwave range using numerical simulation~\cite{Qi2025JAPHelicalCoils}.
	In contrast, the present work focuses on the reflection and absorption of HCC absorber devices at specific microwave frequencies.
	We employ an impedance-matching approach to optimize the absorber performance.
	
	Unlike conventional vertically aligned carbon nanotube (CNT) arrays, HCCs possess a distinct three dimensional helical architecture. This geometry induces a long propagation path, inherent anisotropy, and potential circular dichroism (CD) effects under circularly polarized microwave excitation.~\cite{Qi2025JAPHelicalCoils, Huang2022AdvOptMater, Zhang2025NMLett} Bioinspired helical carbon fibers have been shown to generate chiral asymmetric polarizations that induce broadband microwave absorption and multispectral photonic manipulation.~\cite{Huang2022AdvOptMater} Recent advances have further expanded the design space through hierarchical core shell magnetic nanocomposites,~\cite{Zhao2021Carbon} carbon nanocoil bridged metal organic frameworks,~\cite{Zuo2023Small} and chiral dielectric magnetic trinity foams that achieve ultrabroad effective absorption bandwidths exceeding 14 GHz.~\cite{Zhang2025NMLett} Controllable preparation of helical carbon nanofibers by combustion methods has also yielded minimum reflection losses below -50 dB.~\cite{Guo2024ACSNano}
	
	The geometric parameters of HCCs, including pitch, coil diameter, and array period, strongly govern their microwave response in the 2 to 18 GHz range.~\cite{Sun2025CarbonHelix, Zuo2024AdvFuncMater, Guo2024Small} Qi \textit{et al.} proposed a heuristic resonance relation $f_{\mathrm{res}} = c / (n_{\mathrm{eff}} p)$, with $p$ the coarse grained pitch and $n_{\mathrm{eff}}$ the effective refractive index.~\cite{Qi2025JAPHelicalCoils} Yet this treatment assumes a homogeneous effective medium and ideal impedance matching, overlooking the interfacial discontinuity at the HCC substrate boundary and the anchoring layer mediated gradient matching that we address here.~\cite{Wang2021JCIS, Hou2024Carbon, Liu2023MaterChemPhys}
	
	In this work, we examine the HCC array as a microwave absorber through a two-step analysis. First, we compute the reflection $R^{(0)}$ and absorption $A^{(0)}$ for a bare HCC layer deposited directly on a quartz substrate, with no discernible anchoring layer. We use the heuristic relation $f_{\mathrm{res}} = c / (n_{\mathrm{eff}} p)$ to select representative incident frequencies. Second, we introduce an intermediate anchoring layer of thickness $d_a$ and demonstrate that its variation enables active tuning of impedance matching, yielding absorption maxima at selected frequencies. The anchoring layer thickness thus emerges as a practical control parameter for device optimization.
	
	In the present calculations, we employ a diagonal constitutive tensor and neglect the chiral parameter $\kappa$, as the incident waves are treated as linearly polarized and the absorber performance is governed by dielectric and magnetic losses rather than circular dichroism. Nevertheless, the coarse grained electrodynamic model retains the structural flexibility to incorporate $\kappa$ through off-diagonal terms in the constitutive relations, which would yield distinct refractive indices for left- and right-handed circularly polarized waves. This extension, relevant to the circular dichroism effects reported by Qi et al.~\cite{Qi2025JAPHelicalCoils}, is discussed in Section of Discussions as a natural generalization of the current framework.
	
	The novelty of the present study lies in combining a coarse grained electrodynamic model of HCCs with substrate-aware transmission-line calculations. This allows systematic exploration of absorber performance, impedance matching, and dissipation without relying on full scale electromagnetic simulations. The approach bridges the gap between experimental studies of HCC arrays and practical design rules for microwave absorbers, and provides quantitative insights for the optimization of anchoring layer thickness as a control parameter for absorption enhancement.
	
	\section{Model}
	\label{sec:system}
	
\begin{figure*}[t]
		\centering
		\includegraphics[width=0.96\textwidth]{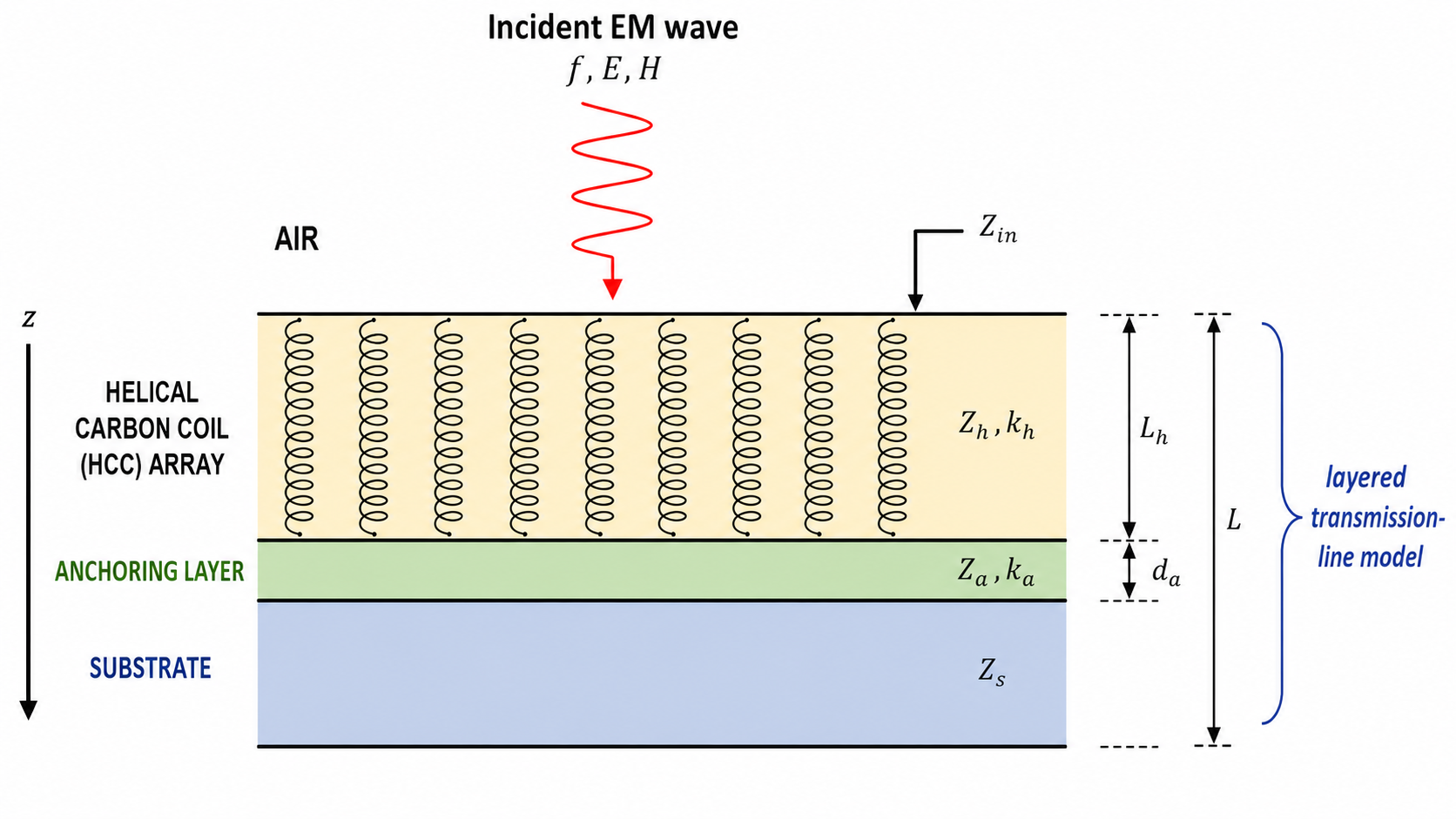}
		\caption{
			Schematic illustration of the coarse grained helical carbon coil (HCC) absorber architecture used in the present work. The device is modeled as a layered structure composed of air, an HCC active layer, an anchoring layer, and a substrate. The incident electromagnetic wave is characterized by frequency \(f\) and field components \(\mathbf{E}\) and \(\mathbf{H}\). The air-side input impedance is denoted by \(Z_{\mathrm{in}}\), while the effective parameters of the HCC layer, anchoring layer, and substrate are represented by \((Z_h,k_h)\), \((Z_a,k_a)\), and \(Z_s\), respectively. The helical active-layer thickness is \(L_h\), the anchoring-layer thickness is \(d_a\), and the total device thickness is \(L\). This schematic motivates the substrate-aware layered transmission-line model used throughout the manuscript to analyze impedance matching, internal attenuation, and absorber performance.
		}
		\label{fig:device_schematic}
	\end{figure*}

	We consider a substrate supported array of vertically aligned HCCs immersed in air. Each helical element is characterized by a pitch $p$, helix radius $R_h$, and total height $L$. The substrate occupies the region \(z<0\), the helical array extends over \(0<z<L\), and air fills the half-space \(z>L\), as shown in Fig.~\ref{fig:device_schematic}. At the interface betweem HCC and substrate, the helical elements are embedded in an anchoring layer of thickness $d_a$. Its electromagnetic properties differ from the main helical section due to denser packing, residual catalyst, or stronger substrate coupling.
	
	The schematic in Fig.~\ref{fig:device_schematic} depicts the coarse grained absorber architecture adopted here. The device is modeled as a layered electromagnetic structure composed of air, a HCC active layer, a thin anchoring layer, and a supporting substrate. The HCC layer is treated as the main dissipative medium, characterized by an effective impedance $Z_h$ and propagation constant $k_h$, while the anchoring region is assigned its own effective parameters ($Z_a$, $k_a$) to account for denser packing, base fixation, and local substrate coupling. The substrate enters through its terminal impedance $Z_s$, and the incident field sees the full stack through the air side input impedance $Z_{\text{in}}$. This layered transmission-line perspective is physically well motivated for the coarse grained modeling of HCC architectures. Recent simulation work on chiral helical carbon coil arrays in the $2$-$18\;\mathrm{GHz}$ range has treated these structures as effective microwave media rather than as atomistically resolved nanocoils.~\cite{Qi2025JAPHelicalCoils} Whereas prior studies have relied predominantly on numerical simulations to characterize the electromagnetic response of HCC arrays,~\cite{Qi2025JAPHelicalCoils} the present work pursues a complementary direction by developing an analytical transmission-line framework and evaluating its predictions for representative parameter values. This framework furnishes explicit design relations for impedance matching, internal dissipation, substrate termination, and anchoring layer effects, all cast within a unified absorber design context.
	We then study the absorption properities of the microwave by this HCC based absorber. 
	\begin{equation}
		A^{(0)}\approx (1-R^{(0)})\left(1-e^{-2\alpha L_h}\right),
		\label{eq:Aapprox}
	\end{equation}
	where $R$ is the entrance reflectance and $\alpha$ is the attenuation constant, therefore, even a strongly lossy medium may be a poor absorber if most of the incoming wave is reflected at the surface. This is the reason that we study the impedance matching absorption properties of the present HCC adsorption device working at the microwave window. 
	
	For a energy lossy HCC array, its effective complex refractive index is $\tilde n=n+i\beta$, with dielectric function $\tilde\varepsilon=\varepsilon'+i\varepsilon''$ and $\mu_r\approx1$. In the weakly to moderately lossy regime, standard weak loss relation shows $\beta\approx \varepsilon''/\big(2\sqrt{\varepsilon'}\big)$ and the attenuation constant $\alpha=\omega\beta/c$.~\cite{YehYariv1984,LandauLifshitz1984} Here, the effective $\varepsilon'$ and $\varepsilon''$ are coarse grained, geometry dependent quantities rather than fixed scalar constants. Qi \etal's full-wave simulations of HCC arrays in the $2$-$18$ GHz range demonstrated that the microwave response depends systematically on geometric parameters including pitch, coil diameter, wire diameter, and array period.~\cite{Qi2025JAPHelicalCoils} 
	
	\section{Theory of Impedances}
	
     We neglect the chiral effect here for the studies on absorption and reflection behaviors of the HCC based device, thus adopt the weak-chirality condition. For the present HCC array with the coarse grained refractive index $n_{\text{eff}}$, the air-side input impedance is
		\begin{equation}
			Z_{\mathrm{in}} = Z_h\,\frac{Z_t + i Z_h \tan(k_h L_h)}{Z_h + i Z_t \tan(k_h L_h)},
			\label{eq:Zin_basic}
		\end{equation}
		where $L_h=L-d_a$ is the thickness of the main helical carbon section ($d_a$ is the anchoring layer thickness), and $Z_h$ and $k_h$ denote its effective impedance and wave number, respectively. Equation~\eqref{eq:Zin_basic} is the standard input-impedance relation of classical transmission-line theory for a finite absorbing layer with characteristic impedance $Z_h$, propagation constant $k_h$, and terminal load $Z_t$.~\cite{YehYariv1984,Wang2021JCIS} This transmission-line framework is widely used in the analysis of microwave absorbers and carbon-based electromagnetic wave absorbing materials, where impedance matching to free space is recognized as one of the key conditions for efficient attenuation and absorption.~\cite{Xiong2024Carbon,Zhang2024ComposB,Saeed2024Alloys} In the present HCC based device, such a coarse grained layered description is also consistent with recent microwave-scale modeling of chiral helical carbon-coil arrays,~\cite{Qi2025JAPHelicalCoils} as well as with earlier studies demonstrating microwave absorption by carbon micro coil systems.~\cite{Motojima2003JAP,Motojima2004}
		
		For the terminal load, we model the system as a two layer stack above the substrate, with a main helical section and an anchoring layer of thickness $d_a$. Let the anchoring region have effective impedance $Z_a$ and wave number $k_a$, while the substrate has impedance $Z_s$. The effective termination seen by the main helical section is then
		\begin{equation}
			Z_t = Z_a\,\frac{Z_s + i Z_a \tan(k_a d_a)}{Z_a + i Z_s \tan(k_a d_a)}.
			\label{eq:Za}
		\end{equation}
		The overall reflection coefficient at the air side is
		\begin{equation}
			r=\frac{Z_{\mathrm{in}}-Z_0}{Z_{\mathrm{in}}+Z_0},
			\;
			R=|r|^2.
			\label{eq:reflectance}
		\end{equation}
		Both the HCC layer and the anchoring layer contribute to the energy lossy of microwave. The imaginary part of the refractive index of HCC array and anchoring layer $\beta_h$ and $\beta_a$, leads to the attenuation constant $\alpha_{h,a} = \omega\beta_{h,a}/c$, then a coarse grained approximation for the transmitted fraction is
		\begin{equation}
			T \approx (1-R)e^{-2(\alpha_hL_h+\alpha_a d_a)},
			\label{eq:Tfull}
		\end{equation}
		and the absorbed fraction becomes
		\begin{equation}
			A = 1-R-T
			\approx
			(1-R)\Bigl[1-e^{-2(\alpha_hL_h+\alpha_a d_a)}\Bigr].
			\label{eq:Afull}
		\end{equation}
		It is worth noting that Eq.~\eqref{eq:Afull} should be understood as a broadband coarse grained approximation rather than an exact multilayer identity, since the exact four layer problem contains internal reflections and interference. It is nevertheless useful for absorber design when the main interest is the overall matching/attenuation trend rather than narrow Fabry-P\'erot features.~\cite{Xiong2024Carbon,Zhang2024ComposB}
		
	\section{Numerical Results of Resonance Guided Absorption of HCC Based Microwave Absorbers}
	\label{sec:predictions}
	
\begin{figure}[htbp]
		\includegraphics[width=0.9\textwidth]{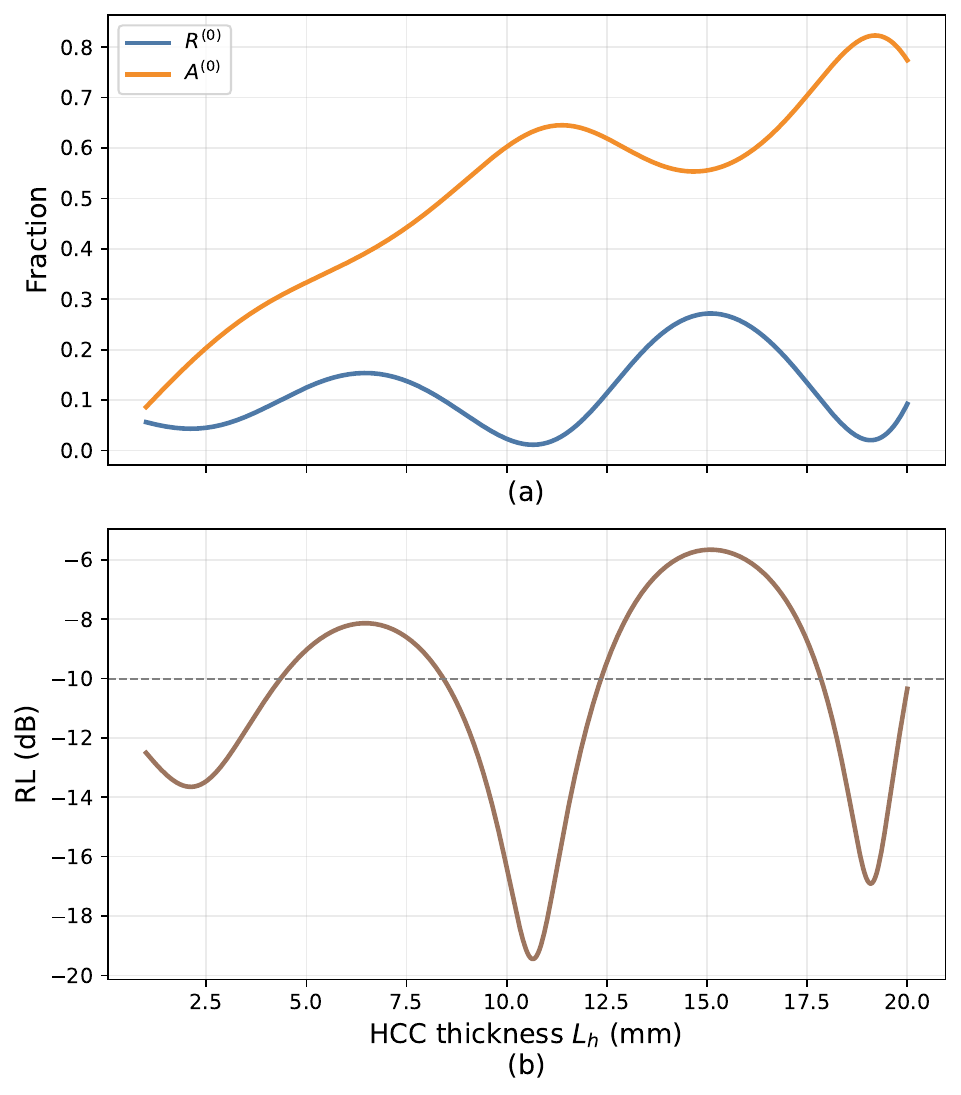}
		\caption{The refraction and absorption of a HCC$\mid$\ch{SiO2} substrate device for an incoming microwave with frequency about $10\,\un{GHz}$. (a) The refraction and absorption fractions denoted by $R^{(0)}$ and $A^{(0)}$; (b) The refection loss of the two layer HCC based absorption device for the $10\,\un{GHz}$ incoming microwave.}
		\label{fig2}
	\end{figure}

	We numerically evaluate the performance of a millimeter scaled HCC array based microwave absorber, in the $2$-$18\,\un{GHz}$ microwave window. We use typical parameters of an ideal HCC device, with the helical pitch of $p=15\,\un{mm}$, and effective refractive index $n_{\text{eff}}=2$, thus the resonant frequency is $f_{\text{res}}=10\,\un{GHz}$, according to the heuristic relationship $f_{\text{res}}=c/\big(n_{\text{eff}}p\big)$. The HCC array has the dielectric constant about $\varepsilon_{r,h}\approx 3.0+0.8i$, thus the loss angle satisfies $\tan\delta\approx 0.3$, and permittivity $\mu_{r,h}\approx 1$, which is in good agreement with recent researches on the electromagnetic response of HCC.~\cite{Qi2025JAPHelicalCoils}, which established that millimeter scale HCC architectures can be treated as effective microwave structures with geometry-dependent response. A typical silicon substrate has $\varepsilon_{r,s}=3.2$ and $\mu_{r,h}\approx 1$. For a ideal two layer system of HCC$\mid$substrate, and the incoming microwave with frequency about $10\,\un{GHz}$, the HCC length dependence of the reflective and absorption fraction, as well as the reflection loss RL in units of dB are shown in Fig.~\ref{fig2}. Here, the superscript $``(0)"$ denotes the quantities associated with this simple two layer HCC device, consisting of HCC array and silicon oxide substrate, thus no anchoring layer. As shown in Fig.~\ref{fig2}, with the HCC array thickness varies from $2$ to $20\,\un{mm}$, the reflection loss varies between about $-20$ to $-6\,\un{dB}$, particularly, the reflection loss approaches its maximum at $L_h\approx 10\,\un{mm}$.	
	
\begin{figure}[htbp]
		\includegraphics[width=0.9\textwidth]{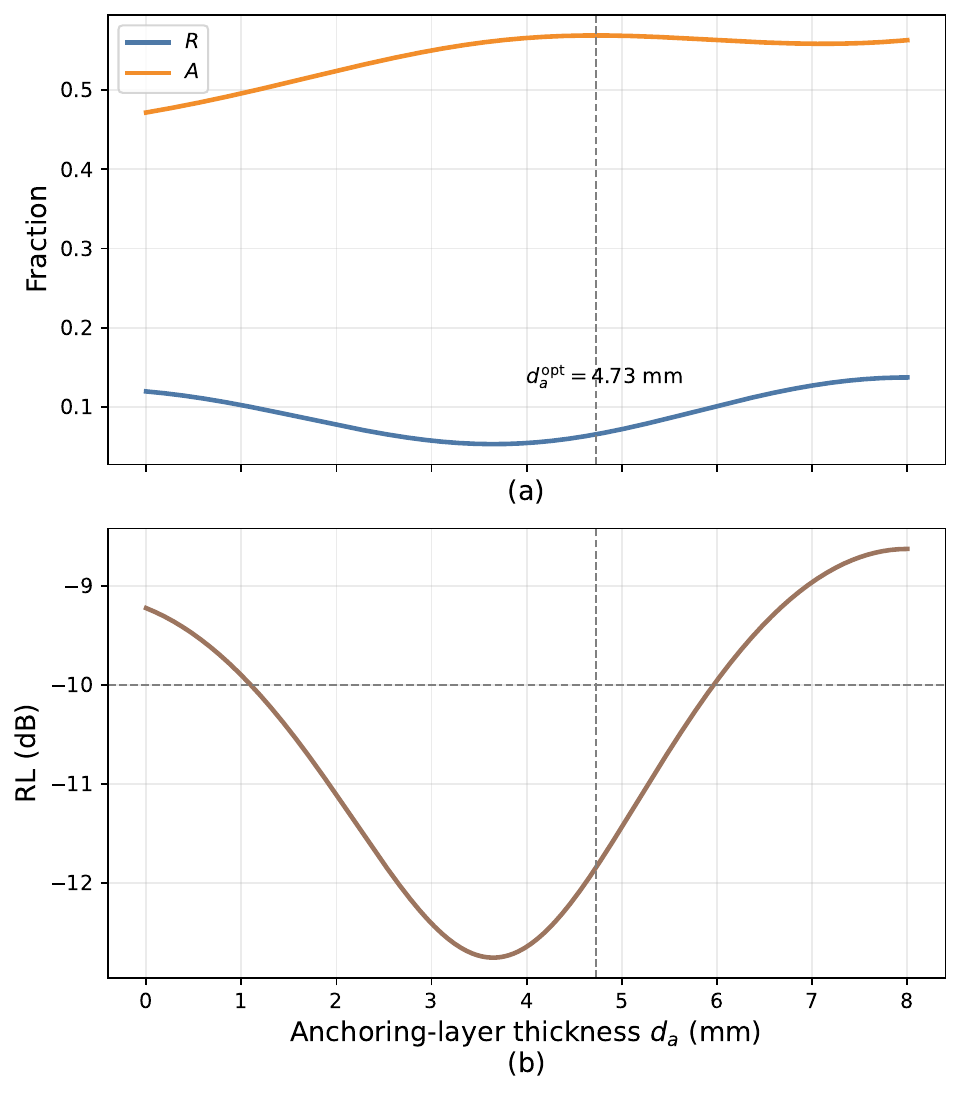}
		\caption{The refraction and absorption of a ``HCC$\mid$anchoring layer$\mid$\ch{SiO2} substrate" sandwich-structured device for an incoming microwave with frequency about $10\,\un{GHz}$. The HCC thickness is set to be $L_h=8\,\un{mm}$. (a) The refraction and absorption fractions denoted by $R$ and $A$. The optimal anchoring layer thickness is labeled for the maximal absorption; (b) The refection loss of the this three layer HCC based absorption device for the $10\,\un{GHz}$ incoming microwave.}
		\label{fig3}
	\end{figure}

	The absorption fraction can be modulated by the millimeter scaled anchoring layer thickness $d_a$, as shown in Fig.~\ref{fig3}, with the HCC thickness is set to $8\,\un{mm}$. The anchoring layer has the parameters of $\varepsilon_{r,a}=3.0+0.3i$ and $\mu_{r,h}\approx 1$. The $d_a$ ranges from $0$ to $8\,\un{mm}$, and numerical results suggest the optimal one $d_{a,\text{opt}}\approx 4.7\,\un{mm}$, corresponding to the maximal absorption. This example suggests that the anchoring layer can modulate the absorption fraction of incoming microwave by the HCC based device with the anchoring layer, manifesting the effects of the anchoring layer as the active modulation medium for the absorption of microwave.   
	
	\section{Discussions}
	\label{sec:model}
	
    The present study focuses on the reflection and absorption characteristics of HCC absorber devices. To use the simple model to illustrate basic physics, we consider a monochromatic microwave incidence, and neglecting the chiral effect in previous sections of Theory and Results.  
    Here, we will discuss the chiral effect due to the geometry of a HCC. The refractive index of a Helical carbon coil is chiral dependent due to its geometry. For an HCC architecture, the local tangent vector of the helical centerline plays an important role for the interaction between electromagnetic and HCC. A circular helix with radius $R_h$ and pitch $p$, has the parameterized description,
	\begin{equation}
		\mathbf r(\phi)=\big(R_h\cos\phi,\;R_h\sin\phi,\;h\phi\big),
		\quad h=\frac{p}{2\pi},
		\label{eq:rhelix}
	\end{equation}
	and its unit tangent vector is,
	\begin{equation}
		\mathbf t(\phi)=\frac{1}{\sqrt{R_h^2+h^2}}\big(-R_h\sin\phi,\;R_h\cos\phi,\;h\big).
		\label{eq:tangent}
	\end{equation}
	For the continuum theory of medium, the anisotropy dielectric function is assumed to be depended on the tangent field $\mathbf t(\phi)$,
	\begin{equation}
		\varepsilon_{ij}(\phi)=\varepsilon_{\perp}\delta_{ij}+\Delta\varepsilon\,t_i(\phi)t_j(\phi),
		\label{eq:epsij}
	\end{equation}
	with $\Delta\varepsilon=\varepsilon_{\parallel}-\varepsilon_{\perp}$. These equations~\eqref{eq:epsij} has used the standard continuum form of a uniaxial anisotropic medium, with the local symmetry axis identified as the tangent field of the helical HCC.\cite{YehYariv1984,LandauLifshitz1984,Qi2025JAPHelicalCoils} Within this coarse grained framework, the isotropic term captures the bulk electromagnetic response, whereas the anisotropic term aligned with the local tangent $\mathbf t(\phi)$  of the helix accounts for the direction dependent transport and polarization effects inherent to the coiled carbon structure. The model is physically reasonable for HCC systems, as experimentally reported carbon coils exhibit substantial electrical conductivity and structure dependent transport behavior.~\cite{Motojima2004,Deng2018PCCP} The present model therefore extends the classical uniaxial medium description to a helical carbon architecture by replacing a fixed optic or transport axis with the local tangent vector $\mathbf{t}(\phi)$ of the coarse grained HCC filament. It is worth noting, HCC arrays are treated as an effective microwave media governed by geometric parameters rather than as atomistically resolved single filaments.~\cite{Qi2025JAPHelicalCoils}
	
	When the electromagnetic wavelength greatly exceeds the helix radius but remains comparable to the pitch, the anisotropy with a spatially rotating principal axis can be coarse grained into an effective reciprocal chiral response,
	\begin{equation}
		\mathbf{D}=\varepsilon_{0}\varepsilon_{r}\mathbf{E}
		-i\xi\,\mathbf{H},
		\qquad
		\mathbf{B}=\mu_{0}\mu_{r}\mathbf{H}
		+i\xi\,\mathbf{E},
		\label{eq:chiralconstit}
	\end{equation}
	where $\xi=\kappa\sqrt{\varepsilon_{0}\mu_{0}}$ is the 
	magnetoelectric coupling parameter, with $\kappa$ the 
	dimensionless chiral parameter induced by the helical 
	geometry. The quantities $\varepsilon_{r}$ and $\mu_{r}$ denote 
	the effective relative permittivity and permeability after 
	homogenization. The constitution equation shown in Eq.~\eqref{eq:chiralconstit} is the 
	standard reciprocal chiral (Pasteur) constitutive relation for 
	an effectively homogeneous bi-isotropic medium, routinely 
	used in chiral metamaterial 
	characterization.\cite{Bassiri1988JOSA,Zhao2010OpticsExpress}
	For propagation approximately along the helical axis, the circular eigenmodes decouple, yielding the refractive indices,
	\begin{equation}
		n_{\pm}=\sqrt{\varepsilon_{r}\mu_{r}}\pm\kappa,
		\label{eq:npm}
	\end{equation}
	with $\pm$ corresponds to right-handed and left-handed circularly polarized modes, respectively,
	whereas the corresponding wave impedance
	\begin{equation}
		Z_{\pm}=\sqrt{\frac{\mu_{0}\mu_{r}}
			{\varepsilon_{0}\varepsilon_{r}}}
		=\sqrt{\frac{\mu_{r}}{\varepsilon_{r}}}
		Z_{0}.
		\label{eq:Zpm}
	\end{equation}
	Thus, in the simple reciprocal isotropic chiral model, the helical geometry produces circular birefringence through $n_+\neq n_-$, while
	the intrinsic wave impedance remains the same for the two circular eigenmodes. In our previous sections, we do not consider the difference between the left- and right-handed circular polar incoming wave, and the present work can be extended to the including the chiral index, by introducing the phenomenological chiral parameter $\kappa$ for the circular dichroism effect involved phenomena. 
	
	In the study of the effects of anchoring layer, we start with an ideal model, and this layered treatment elucidates the physical roles of the substrate and anchoring layer in the microwave Adsorption design. The substrate sets the terminal impedance $Z_s$, which governs the phase of the reflected wave and thereby tunes the standing wave interference and the frequency response of the effective input impedance $Z_{\text{in}}$. The anchoring layer, acting as an electromagnetic transition between the HCC and the substrate, is critical for achieving impedance matching.  when its effective impedance $Z_a$ and thickness $d_a$ are chosen such that $Z_a \approx \sqrt{Z_h Z_s}$ and the optical thickness $n_a d_a \sim \lambda/4$, it functions as an anti-reflection coating that suppresses interfacial reflection; conversely, if $Z_a$ is strongly mismatched with either $Z_h$ or $Z_s$, or if $d_a$ yields destructive interference, additional reflection is introduced at this interface and the overall microwave adsorption performance degrades. In realistic HCC device, the anchoring layer should therefore be treated as an active electromagnetic functional layer for tuning $Z_{\text{in}}(\omega)$, rather than as a negligible geometric detail.
	
	\section{Conclusion}

	So far, we have presented a theoretical framework for analyzing the microwave absorption performance of substrate supported HCC arrays. Guided by the heuristic relation $f_{\mathrm{res}} = c / (n_{\mathrm{eff}} p)$, we selected representative microwave frequencies in the $2$-$18\,\un{GHz}$ range and investigated both ``HCC$\mid$substrate" and ``HCC$\mid$anchoring layer$\mid$substrate" absorber. 
	
	Our analysis indicates that the HCC$\mid$substrate system alone exhibits absorption for a special HCC thickness, and the reflection at the air interface may lead a limit absorption away from the optimal HCC thickness. By introducing a carbon based anchoring layer of finite thickness between the HCC array and substrate, for a certain HCC thickness away from the optimal one, the effective input impedance can be tuned, resulting in enhanced absorption at the selected frequency. This establishes the anchoring layer as an effective active parameter for optimizing device performance. 
	
	Additionally, our quasi quantitative analysis does not consider the chiral parameter $\kappa$ contributes negligibly to absorption in this frequency band, and use an effective isotropic refractive index $n_{\mathrm{eff}}$ for design purposes, instead. The present framework provides practical guidance for future experiments and device fabrication, offering a physically grounded route to design high performance HCC-based microwave absorbers. By introducing the phenomenological chiral parameter $\kappa$ into the electromagnetic constitutive relationship, the present model can be extended to study possible circular dichroism behavior of the HCC based device as a microwave absorber, in a straightforward way.  
	
	Overall, this study combines the structural features of HCCs, substrate interactions, and impedance matching principles to deliver a predictive model with direct applicability to engineering microwave absorbers with enhanced attenuation and reflection control.
	
	\end{CJK}
	

	\bibliographystyle{aipnum4-2}
	\bibliography{H}
	
\end{document}